\documentclass[12pt]{article}\usepackage{quanfruit}
\begin{document}

\title{Java Application that Outputs\\
Quantum Circuit for Some\\
NAND Formula Evaluators}

\author{Robert R. Tucci\\
        P.O. Box 226\\
        Bedford,  MA   01730\\
        tucci@ar-tiste.com}

\date{ \today}

\maketitle

\vskip2cm
\section*{Abstract}
This paper introduces
QuanFruit v1.1,
a Java application
available for free. (Source code
included in the distribution.)
Recently, Farhi-Goldstone-Gutmann (FGG)
wrote a paper arXiv:quant-ph/0702144
that proposes a quantum algorithm for
evaluating NAND formulas.
QuanFruit outputs a quantum
circuit for the FFG algorithm.

\section{Introduction}

This paper introduces
QuanFruit v1.1,
a Java application
available\cite{QuanSuite} for free. (Source code
included in the distribution.)
Recently, Farhi-Goldstone-Gutmann (FGG)
wrote a paper\cite{FGG07}
that proposes a quantum algorithm for
evaluating NAND formulas.
QuanFruit outputs a quantum
circuit for the FFG algorithm.

We say a unitary operator
acting on a set of qubits has been
compiled if
it has been expressed
as a SEO (sequence of elementary
operations, like CNOTs and single-qubit operations).
SEO's are often represented as quantum circuits.

There exist software
(quantum compilers)
like Qubiter\cite{Tuc99}
for compiling arbitrary unitary
operators (operators that have no a priori
known structure).
QuanFruit is
a special purpose quantum
 compiler.
It is special purpose
in the sense that it can only
compile unitary operators
that have a very definite, special
structure.

The QuanFruit
application is part
of a suite of Java applications called QuanSuite.
QuanSuite applications are all
based on a common class library called QWalk.
Each QuanSuite application compiles a
different kind of quantum evolution operator.
The applications
output a quantum circuit
that equals the input
evolution operator.
We have introduced 6 other
QuanSuite applications in
2 earlier papers.
Ref.\cite{qtree} introduced QuanTree
and QuanLin.
Ref.\cite{qfou} introduced
QuanFou, QuanGlue, QuanOracle,
and QuanShi.
QuanFruit
calls methods from these 6
previous applications, so
it may be viewed
as a composite of them.

Before reading this paper,
the reader should read
Refs.\cite{qtree} and \cite{qfou}.
Many
explanations in
Refs.\cite{qtree} and \cite{qfou}
still apply to this paper.
Rather than repeating
such explanations in this paper,
the reader will be
frequently referred to Refs.\cite{qtree} and \cite{qfou}.

The goal of all
QuanSuite applications, including QuanFruit, is
to compile an input evolution operator $U$.
$U$ can
be specified either directly
(e.g. in QuanFou, QuanShi),
or by giving a Hamiltonian $H$ such
that $U = e^{iH}$ (e.g. in QuanGlue and QuanOracle).

The standard definition of
the evolution operator
in Quantum Mechanics is
$U= e^{-itH}$, where
$t$ is time and $H$
is a Hamiltonian. Throughout
this paper, we will set
$t = -1$ so $U = e^{iH}$.
If $H$ is proportional
to a coupling constant $g$,
reference to time can be
 restored easily by
replacing the symbol $g$ by
$-tg$, and the symbol $H$ by $-tH$.

\section{Input Evolution Operator}

The input evolution operator for QuanFruit is
$U_{fruit}=e^{iH_{fruit}}$, where

\beq
H_{fruit} =
\left[
\begin{array}{c|c|c|c|c|c|c}
\multicolumn{4}{c|}{}&h^\dagger_{glue}&\;\;\;\;\;&\\ \cline{5-7}
\multicolumn{4}{c|}{}&&&\\ \cline{5-7}
\multicolumn{4}{c|}{}&&&\\ \cline{5-7}
\multicolumn{4}{c|}{\rb{4ex}{$H_{line}$}}&&&\\ \hline
h_{glue}&&&&\multicolumn{2}{c|}{}&\\ \cline{1-4}\cline{7-7}
&&&&\multicolumn{2}{c|}{\rb{1ex}{$H_{tree}$}}&h_{ora}^\dagger\\ \hline
\;\;\;\;\;&\;\;\;\;\;&\;\;\;\;\;&\;\;\;\;\;&&h_{ora}&
\end{array}
\right]
\;.
\eeq

\begin{figure}[h]
    \begin{center}
    \includegraphics[width=4in]{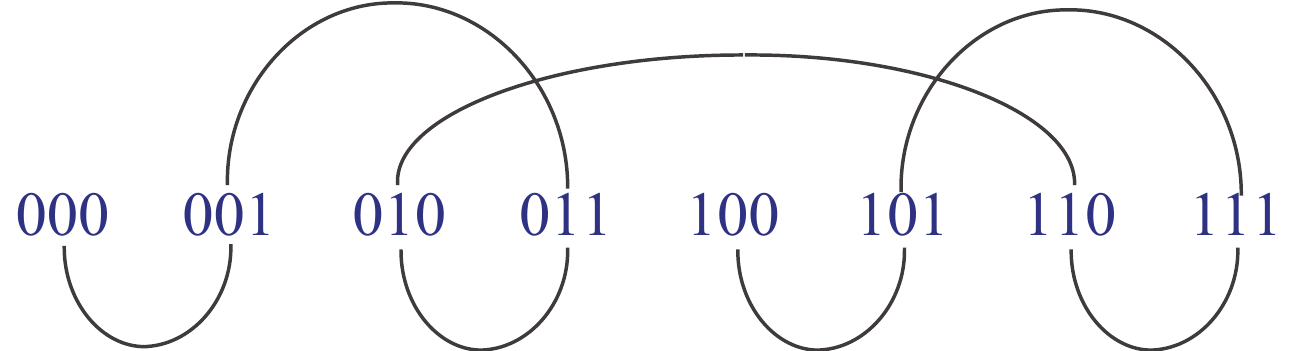}
    \caption{Line (open string) with 8 nodes
    }
    \label{fig-line-graph}
    \end{center}
\end{figure}

$H_{line}\in\RR^{N_{S,line}\times N_{S,line}}$
where $N_{S,line}=2^{N_{B,line}}$ for some
positive integer $N_{B,line}$.
$H_{line}$ is proportional to the
incidence matrix for a line graph,
where the edges of the graph
connect states
that are consecutive in a Gray order.
For example, for $N_{B,line}=3$,
the graph of Fig.\ref{fig-line-graph}
yields:
\beq
H_{line} = g
\begin{array}{c|c|c|c|c|c|c|c|c|}
&\p{000}&\p{001}&\p{010}&\p{011}&\p{100}&\p{101}&\p{110}&\p{111} \\ \hline
\p{000}&0&1& & & & & &  \\ \hline
\p{001}&1&0& &1& & & &  \\ \hline
\p{010}& & &0&1& & &1&  \\ \hline
\p{011}& &1&1&0& & & &  \\ \hline
\p{100}& & & & &0&1& &  \\ \hline
\p{101}& & & & &1&0& &1  \\ \hline
\p{110}& & &1& & & &0&1  \\ \hline
\p{111}& & & & & &1&1&0 \\ \hline
\end{array}
\;,
\label{eq-h-line}
\eeq
where $g$ is a real number that we will call
the {\bf coupling constant}.

\begin{figure}[h]
    \begin{center}
    \includegraphics[height=1.5in]{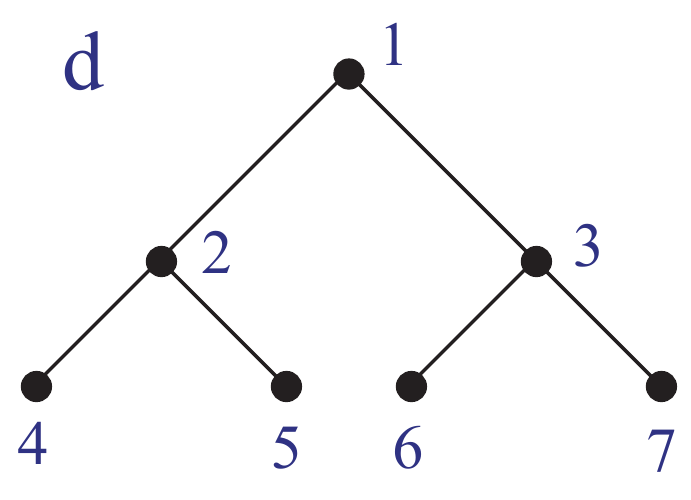}
    \caption{Binary tree with 8 nodes
    }
    \label{fig-tree-graph}
    \end{center}
\end{figure}

$H_{tree}\in\RR^{N_{S,tree}\times N_{S,tree}}$
where $N_{S,tree}=2^{N_{B,tree}}$ for some
positive integer $N_{B,tree}$.
$H_{tree}$ is proportional to the
incidence matrix for a
balanced-binary tree graph.
For example, for $N_{B,tree}=3$,
the graph of Fig.\ref{fig-tree-graph}
yields:
\beq
H_{tree} = g
\begin{array}{c|c|c|c|c|c|c|c|c|}
&\p{d}&\p{1}&\p{2}&\p{3}&\p{4}&\p{5}&\p{6}&\p{7} \\ \hline
\p{d}& & & & & & & &  \\ \hline
\p{1}& & &1&1& & & &  \\ \hline
\p{2}& &1& & &1&1& &  \\ \hline
\p{3}& &1& & & & &1&1  \\ \hline
\p{4}& & &1& & & & &  \\ \hline
\p{5}& & &1& & & & &  \\ \hline
\p{6}& & & &1& & & &  \\ \hline
\p{7}& & & &1& & & &  \\ \hline
\end{array}
\;,
\label{eq-h-tr}
\eeq
where $g$ is the same
coupling constant as before.

$h_{glue}\in \RR^{1\times 1}$. In fact,

\beq
h_{glue} = g\ket{N_{S,line}+1}\bra{d}
\;.
\eeq
Here $\ket{N_{S,line}+1}$
 labels the
god state of the tree, the one
with children but no parents.(
$N_{S,line}$ labels the dud node)
We will call
$d\in Z_{0,N_{S,line}-1}$
the {\bf line door}.
If $d=0$, then the tree is connected
to a tail of states. For $N_{B,line}=3$,
from  Fig.\ref{fig-line-graph}, if $d=2$, then the
tree is connected to the midpoint of the
line of states (``runway").

The number of
leaves in the tree is half the number of
nodes in the tree:
$N_{S,lvs}=\frac{N_{S,tree}}{2}$.
Also, $N_{S,lvs}=2^{N_{B,lvs}}$
for some positive integer $N_{B,lvs}$.
$h_{oracle}\in \RR^{N_{S,lvs}\times N_{S,lvs}}$.
In fact,

\beq
h_{oracle}
= g
\left[
\begin{array}{cccc}
x_0&&&\\
&x_1&&\\
&&\ddots&\\
&&&x_{\nlvs-1}
\end{array}
\right]
\;,
\eeq
where $x_k\in Bool$ are
the inputs to the NAND formula.

The dimension of the matrix $H_{fruit}$ is
not generally a power of two.
To represent it as a quantum circuit,
we need to extend it to
$diag(H_{fruit},0)\in \RR^{\ns\times\ns}$,
where
\beq
\ns = 2^\nb\;,\;\;
\nb = \min\{N: N_{S,line} +
\frac{3}{2}N_{S,tree}\leq 2^N\}
\;.
\label{eq-nb}
\eeq

Define
\beq
H_{glue} = h_{glue} + h_{glue}^\dagger\;,\;\;
H_{ora} = h_{ora} + h_{ora}^\dagger
\;.
\label{eq-htoH}
\eeq
(This last equation is fine as an
operator statement, but as a matrix
statement, $h_{glue}$ and $h_{glue}^\dagger$
must be ``padded" with zeros to make the
equation true. By ``padding a matrix
with zeros'', we mean
embedding it in a larger matrix,
the new entries
being zeros.)

One can split $H_{fruit}$
into two parts, which we
call the
{\bf bulk Hamiltonian} $H_{bulk}$
and the {\bf boundary corrections Hamiltonian}
$H_{corr}$:

\beq
H_{fruit} = H_{bulk} + H_{corr}
\;,
\eeq
where

\beq
H_{bulk} = H_{line} + H_{tree}
\;\;\;,\;\;\;
H_{corr} = H_{glue} + H_{ora}
\;.
\eeq
(Again,
this last equation requires
zero padding if considered a matrix equation.)
Note that $[H_{line},H_{tree}]=0$
and $[H_{glue},H_{ora}]=0$.

For $r=1,2,3,\ldots$,
if $U = L_r(g) + \calo(g^{r+1})$,
we say
$L_r(g)$ {\bf approximates
(or is an approximant) of order}
$r$
for  $U$.

Given an approximant
$L_r(g) + \calo(g^{r+1})$
of $U$,
and some $\nt= 1, 2, 3\,\ldots$,
one can approximate $U$ by
$\left(L_r(\frac{g}{\nt})\right)^\nt
+ \calo(\frac{g^{r+1}}{\nt^r})$.
We will refer to this as {\bf Trotter's trick},
and to $\nt$ as the {\bf number of trots}.

For $N_{T,line}=1$, QuanFruit
approximates $e^{iH_{line}}$
with a Suzuki approximant of
order $r_{line}=2, 4, 6, \ldots$
that is derived
in Ref.\cite{Theory}.
QuanFruit also applies
the Trotter trick with
$N_{T,line}>1$ trots to
the $N_{T,line}=1$ approximant of
$e^{iH_{line}}$.

For $N_{T,tree}=1$,
QuanFruit always approximates
$e^{iH_{tree}}$ with
an approximant of order 3,
that is derived
in Ref.\cite{Theory}.
QuanFruit also applies
the Trotter trick with
$N_{T,tree}>1$ trots to
the $N_{T,tree}=1$ approximant of
$e^{iH_{tree}}$.

Ref.\cite{Theory}
gives exact (to numerical precision)
compilations of the glue and oracle
parts of $U_{fruit}$.
QuanFruit uses these compilations, so
 the Order of
the Suzuki (or other) Approximant
and the Number of Trots
do not arise in QuanFruit,
for either the glue or the oracle.

For $N_{T, meta}=1$, QuanFruit
also approximates $e^{iH_{fruit}}$
with a Suzuki approximant of
order $r_{meta}=2, 4, 6, \ldots$.
Recall that $S_2(t) = e^{A\frac{t}{2}}
e^{Bt}e^{A\frac{t}{2}}$
for $t\in\RR$
is the second order Suzuki approximant,
and higher order ones are defined
recursively from this one.
Thus, all Suzuki approximants are specified
by giving two functions of $t$,
$e^{At}$ and $e^{Bt}$.
To get a ``meta" Suzuki approximant,
we  set $e^{At} = e^{i(H_{bulk})_{g\rarrow t}}$
and
$e^{Bt} = e^{i(H_{corr})_{g\rarrow t}}$.
QuanFruit also applies
the Trotter trick with
$N_{T,meta}>1$ trots to
the $N_{T,meta}=1$ approximant of
$e^{iH_{fruit}}$.

\section{The Control Panel}

\begin{figure}[h]
\begin{center}
    \includegraphics[height=7in]{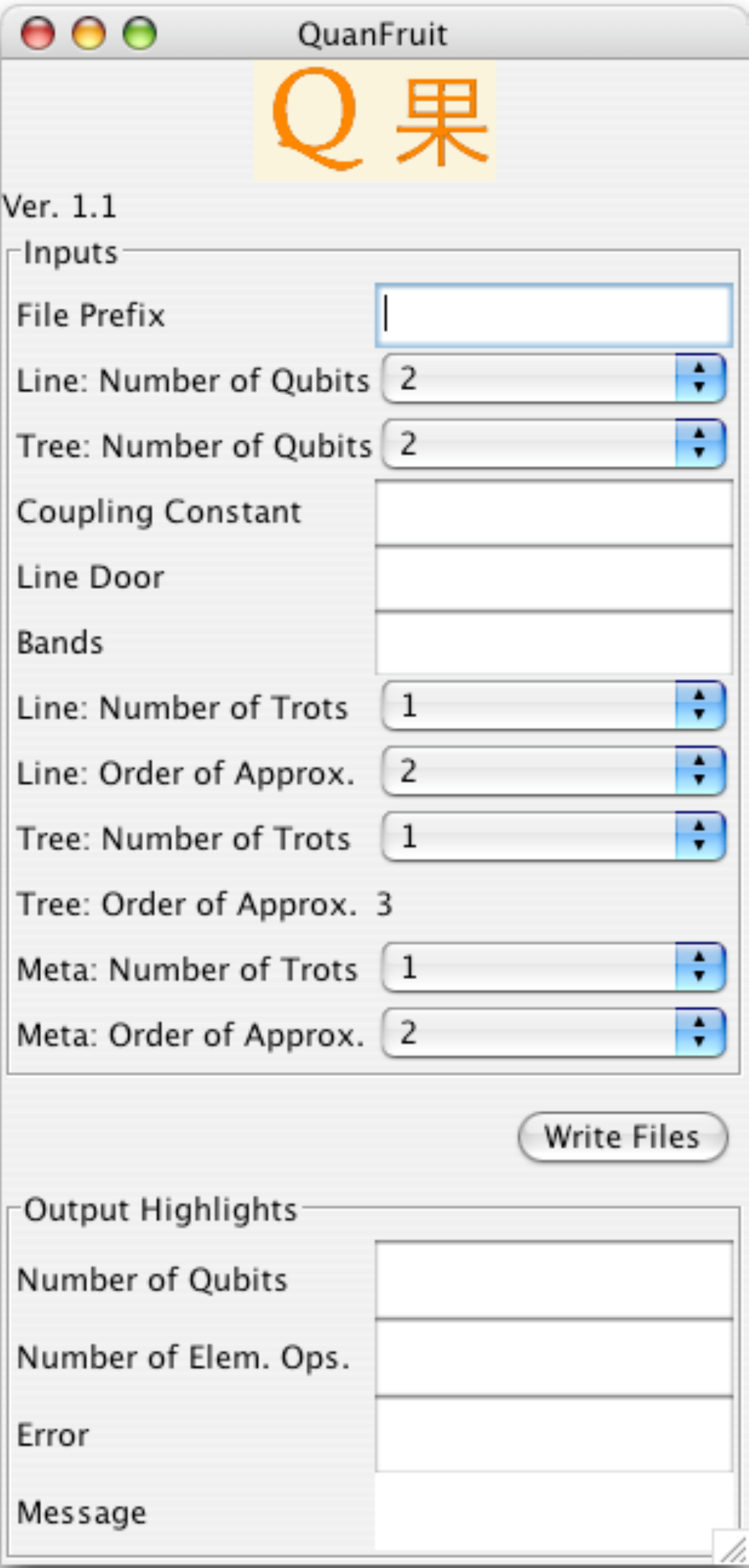}
    \caption{Control Panel of QuanFruit}
    \label{fig-qfruit-main}
\end{center}
\end{figure}

Fig.\ref{fig-qfruit-main} shows the
{\bf Control Panel} for QuanFruit. This is the
main and only window of the application. This
window is
open if and only if the application is running.

The Control Panel
allows you to enter the following inputs:
\begin{description}

\item[File Prefix:] Prefix to the
3 output files
that are written when you press the
{\bf Write Files} button. For example,
if you insert {\tt test} in this text field,
the following 3 files will be written:
\begin{itemize}
\item
{\tt test\_qfru\_log.txt}
\item
{\tt test\_qfru\_eng.txt}
\item
{\tt test\_qfru\_pic.txt}
\end{itemize}

\item[Line: Number of Qubits:]
The parameter $N_{B,line}$ defined above.

\item[Tree: Number of Qubits:]
The parameter $N_{B,tree}$ defined above.

\item[Coupling Constant:] The parameter
$g\in \RR$ defined above.

\item[Line Door:] The parameter $d\in Z_{0,N_{S,line}-1}$
defined above.

\item[Bands:]You must enter here an even number
of integers separated by any non-integer,
non-white space symbols. Say you enter
$a_1,b_1,a_2,b_2,\ldots ,a_n,b_n$.
If $x_k\in Bool$ for $k\in Z_{0,\nlvs-1}$
are as defined above, then
$x_k=1$ iff $k\in
Z_{a_1,b_1}\cup Z_{a_2,b_2} \ldots \cup Z_{a_n,b_n}$.
Each set $Z_{a_k,b_k}$ is a ``band".
If $a_k=b_k$, the band has a single element.
QuanFruit checks that
$0\leq a_0$, $b_n\leq(\nlvs-1)$, and
$b_k-a_k\geq 0$ for all $k$.
It also checks that
$a_{k+1}-b_k\geq 2$. (If
$a_{k+1}-b_k= 1$, bands $k+1$ and $k$
can be merged. If
$a_{k+1}-b_k= 0, -1, -2, \ldots$,
bands $k+1$ and $k$ overlap.)

\item[Line: Number of Trots:]
The parameter $N_{T,line}$ defined above.
\item[Line: Order of Approximant:]
The parameter $r_{line}$ defined above.

\item[Tree: Number of Trots:]
The parameter $N_{T,tree}$ defined above.
\item[Tree: Order of Approximant:]
This parameter is always 3.

\item[Meta: Number of Trots:]
The parameter $N_{T,meta}$ defined above.
\item[Meta: Order of Approximant:]
The parameter $r_{meta}$ defined above.

\end{description}

The Control Panel displays the
 following
outputs:
\begin{description}

\item[Number of Qubits:]
The parameter $\nb$ defined by Eq.(\ref{eq-nb}).

\item[Number of Elementary Operations:]
The number of elementary operations
in the output quantum circuit.
If there are no LOOPs, this is
the number of lines in the English File, which
equals the number of lines in the
Picture File.
When there are LOOPs, the
``{\tt LOOP k REPS:$\nt$}" and
``{\tt NEXT k}" lines are not counted,
whereas the lines between
``{\tt LOOP k REPS:$\nt$}" and
``{\tt NEXT k}"
are counted $\nt$ times.

\item[Error:] The distance in the
Frobenius norm between the input evolution
operator and the output
quantum circuit (i.e., the SEO given in
the English File).
For a nice review of matrix norms, see
Ref.\cite{Golub}.
For any
matrix $A\in\CC^{n\times n}$, its Frobenius
norm is defined as
$\|A\|_F = \sqrt{\sum_{j,k} A_{j,k}A^*_{j,k}}
$.
Another common matrix norm is the 2-norm.
The 2-norm $\|A\|_2$ of $A$
equals the largest singular value of $A$.
The Frobenius and 2-norm of $A$
are related by\cite{Golub}:
$
\|A\|_2 \leq \|A\|_F \leq \sqrt{2}\|A\|_2
$.

\item[Message:]
A message appears in this text field
if you press
{\bf Write Files} with a bad input.
The message tries to explain
the mistake in the input.

\end{description}

\section{Output Files}

Pressing the {\bf Write Files} button
of the Control Panel of QuanFruit generates
3 files (Log, English, Picture).
These files are analogous
to their namesakes for QuanTree,
QuanLin and other
QuanSuite applications.
Ref.\cite{qtree} explains
how to interpret them.

\section{Behind the Scenes: \\
Code Innovations in QuanSuite, QWalk}

The QuanSuite applications,
based on the QWalk class library,
exhibit some
code innovations
that you will find very helpful.
Hopefully,
these innovations will
become commonplace in
future quantum computer software.

\begin{itemize}
\item {\bf QWalk class library does
most of the work in all QuanSuite applications:}
Look in the source folder for any
of the QuanSuite applications.
You'll find that it contains only 3 or
4 classes. Most of the classes
are in the source folder
for QWalk. That's because
most of the work is done by
the QWalk class library,
which is independent of the QuanSuite application.

\item {\bf Reusability of SEO writers:}
Look at the class
{\tt FruitSEO\_writer}
in the source folder
for QuanFruit. You'll find that
{\tt FruitSEO\_writer}
utilizes the methods
{\tt GlueSEO\_writer()},
{\tt OracleSEO\_writer()},
{\tt TreeSEO\_writer()},
{\tt LineSEO\_writer()},
and
{\tt ShiftSEO\_writer()}.
Thus,
{\tt FruitSEO\_writer}
delegates its SEO writing
to methods from
the QuanSuite applications:
QuanGlue, QuanOracle, QuanTree, QuanLin
and
QuanShi.
In fact, QuanFruit can
be viewed as a composite
of these simpler QuanSuite
applications.
This reusability of SEO
writers is made possible
by the novel technique
described in Appendix
\ref{app-pad}.

\item {\bf Nested Loops:}
The English and Picture
files of QuanSuite applications
can have LOOPs within LOOPs.
This makes the
English and Picture
files shorter,
without loss of information.
However, if you want
to multiply
out all the operations
in an English
file (this is what
the class {\tt SEO\_reader}
in QWalk does), then having nested loops
makes this task more difficult.
 {\tt SEO\_reader}
of QWalk
is sophisticated enough
to understand nested loops.

\item {\bf Painless object oriented
implementation of Suzuki approximants
and Trotter's trick:}
Higher order Suzuki approximants
can be implemented painlessly
by using the classes:
{\tt QWalk/src/SuzFunctions}
and {\tt  QWalk/src/SuzWriter}.
See the class
{\tt QuanLin/src/LineSEO\_writer}
for an example of how it's done.
Essentially, all you have to
do is to override the two abstract
methods in
{\tt QWalk/src/SuzFunctions}.

Trotter's trick can also be easily
implemented in a  QuanSuite
application, by
using LOOP and NEXT lines
in the English file. See the
{\tt write()} method
of {\tt QuanLin/src/LineSEO\_writer}
for an example.

\end{itemize}

\appendix
\section{Appendix: $P_0$ Padding and
State Shifting}\label{app-pad}

Suppose
we know how to compile $e^{iH}$.
Is it possible to
use this compilation
to compile $e^{i\;\;diag(Z, H, Z')}$,
where $Z$ and $Z'$ are
square matrices of
zeros? The answer is yes,
as we show next.

Suppose $\hat{N}_S > \ns$
where $\ns = 2^\nb$ and
$\hat{N}_S = 2^{\hat{N}_B}$,
for some positive integers
$\nb$ and $\hat{N}_B$.
Given a Hamiltonian $H$,
define a zero padded version of it called $\hat{H}$:

\beqa
\hat{H}_{\hat{N}_S\times\hat{N}_S} &=&
\left[
\begin{array}{c|c}
H_{\ns\times\ns}&\\
\hline
&0_{(\hat{N}_S-\ns)\times(\hat{N}_S-\ns)}
\end{array}
\right]\\
&=&
\nbar(\hat{N}_B-1)
\otimes
\nbar(\hat{N}_B-2)
\otimes
\cdots
\otimes
\nbar(\nb+1)
\otimes
\nbar(\nb)
\otimes H_{\ns\times\ns}
\;.
\eeqa
As usual, $\nbar() = P_0()$.
We will say that $H$ has been padded
with $P_0$'s to obtain $\hat{H}$.
Now let
$U_{shift}^{(k)}$
be the unitary operation
that shifts state $\ket{x}$ to $\ket{(x+k)\mod
\hat{N}_S}$, with $x,k\in Z_{0,\hat{N}_S-1}$.
The application
QuanShi gives a compilation of
$U_{shift}^{(k)}$.
Using $U_{shift}^{(k)}$,
one
can define a matrix $\hat{H}^{(k)}$
from $\hat{H}$ as follows:

\beqa
(\hat{H}^{(k)})_{\hat{N}_S\times\hat{N}_S} &=&
\left[
\begin{array}{c|c|c}
0_{k\times k}&&\\
\hline
&H_{\ns\times\ns}&\\
\hline
&&0_{(\hat{N}_S-\ns-k)\times(\hat{N}_S-\ns-k)}
\end{array}
\right]\\
&=&
(U_{shift}^{(k)})^{\dagger}
\hat{H}_{\hat{N}_S\times\hat{N}_S}
U_{shift}^{(k)}
\;.
\eeqa

It is now readily apparent
that a SEO for $e^{i\;\;diag(Z, H, Z')}$
can be obtained from a SEO for $e^{iH}$
by $P_0$ padding $H$ and
then state shifting it with
$U_{shift}^{(k)}$.

The compilations
of
$e^{iH_{line}}$ (given in QuanLin),
$e^{iH_{tree}}$ (given in QuanTree) and
$e^{iH_{ora}}$ (given in QuanOracle),
are all utilized by QuanFruit
via this
$P_0$ padding/state-shifting method.

\end{document}